\documentclass[12pt,preprint,aps,showpacs,color,floatfix]{revtex4}
\usepackage{epsfig,color}
\usepackage{amsmath}

\begin{document}
\title{$K^+$ and $K^-$ potentials in hadronic matter are observable quantities}

\author{Aman D. Sood, Ch. Hartnack, and
J. Aichelin }
\address{
$^1$SUBATECH,
Laboratoire de Physique Subatomique et des
Technologies Associ\'ees \\University of Nantes - IN2P3/CNRS - Ecole des Mines
de Nantes \\
4 rue Alfred Kastler, F-44072 Nantes, Cedex 03, France\\}
\begin{abstract}
The comparison of $K^+$ and $K^-$ spectra at  low transverse momentum in light symmetric heavy ion reactions
at energies around 2 AGeV allows for a direct experimental determination of the strength of the $K^+$ as well as of the $K^-$ nucleus potential. Other little known or unknown input quantities like the production or rescattering  cross sections  of $K^+$ and $K^-$ mesons do not spoil this signal.  This result, obtained by simulations of these reactions with the Isospin Quantum Molecular Dynamics (IQMD) model,  may solve the longstanding question of the behaviour of the $K^-$ in hadronic matter and especially whether a $K^-$ condensate can be formed in heavy ion collisions.  
\end{abstract}
\date{\today}

\maketitle
How meson properties change in matter has  theoretically been investigated
since many years. Experimentally measured phase shifts allow via the $t\rho$
approximation to predict the optical potential in matter at low densities.
When higher densities are of interest more complicated approaches have to
be employed and many efforts have been made to investigate the properties
of $\rho, \omega, K^+$ and $K^-$ mesons in matter \cite{Eichstaedt:2007zp,Riek:2010gz,Korpa:2004ae,Tolos:2006ny}.
 Especially the production of strange mesons has created a lot of interest because it has been proposed
to use them to test the properties of the nuclear environment, in particular of the nuclear equation of state \cite{Aichelin:1986ss}. For the $K^+$, which 
cannot form resonances in matter, the nuclear matter calculations agree  \cite{Korpa:2004ae} with those
based on the scattering length. This presents evidence that the calculations are reliable.
For the $K^-$ which forms nuclear resonances, especially the $\Lambda(1405)$, which may melt in matter \cite{Koch:1994mj}, coupled channel calculations have to be employed and the challenge has to 
be met to calculate them selfconsistently. The theoretical predictions launched by different groups differ
substantially \cite{Lutz:2001dq,Ramos:1999ku,Tolos:2006ny} because several of the quantities which enter such calculations, like in-medium coupling constants and the in-medium dressing of the different particles,
are only vaguely known or unknown. Therefore it is highly desirable to identify observables which allow for an 
experimental determination of these quantities.

Simulations show that several observables are sensitive to the in medium properties of mesons. 
The principal problem for extracting  precise information on these properties  is, however,  
 that almost all observables depend simultaneously not only on the $K^-$ nucleus potential but also on several other 
input quantities which are only vaguely known. They include  the lifetime of the $\Delta$ and the modification of $\sigma_{NN \to K^{\pm}X}$ in the medium, the only theoretically known $\sigma_{N\Delta \to K^+N\Lambda} $ cross section and the little known cross sections for the production  of the $K^-$ in secondary interactions $BY \to BB K^-$ or  $\pi Y \to K^- N$ (where $Y \to \Lambda, \Sigma$) which dominate the $K^-$  production in heavy systems\cite{PR}. This new reaction channels (occurring only in HIC ) link the $K^-$ to the $K^+$ production. Thus all the uncertainties related to the production of $K^+$ are inherited by $K^-$. To minimize this influence it is useful to work with ratios of the   $K^+$ and $K^-$ spectra. 

The situation were much better if experiment  would provide an observable which depends on the $K$ potentials only 
and which is not spoiled by other little or unknown quantities.  In this letter we will show that the 
ratio of the $K^+$ and $K^-$  momentum spectra at small momentum in light systems is such an observable.

In order to study this observable and in order to make sure that it does not depend on other input quantities 
we have separated the $K^-$ into 2 classes (by tracing back $K^-$ to its corresponding anti strange partner $K^+$).\\ (a) $K^-$ coming directly from  reactions like $BB \to BB K^+ K^-$ called direct contribution  and abbreviated in the figures by Dir\\
(b) $K^-$ coming from $\pi Y$ or $BY\to K^-$ abbreviated in the figures by Y.\\

The  energy of kaons in medium  $\omega({\bf{k}},\rho)$, is given by \cite{SchaffnerBielich:2000jy,Schaffner:1996kv}:
\begin{equation}
\omega({\bf{k}},\rho)= \sqrt{\left( {\bf{k}-\alpha \Sigma_v} \right)^2+m^2 + m
\alpha \Sigma_s} \pm \alpha\Sigma_v^0 \label{schaf}
\end{equation}
with $\alpha = 1$, with  a scalar self energy $\Sigma_s$ and a vector self energy ($\Sigma_v^0, {\bf \Sigma_v})$,
\begin{figure}[b]
\psfig{figure=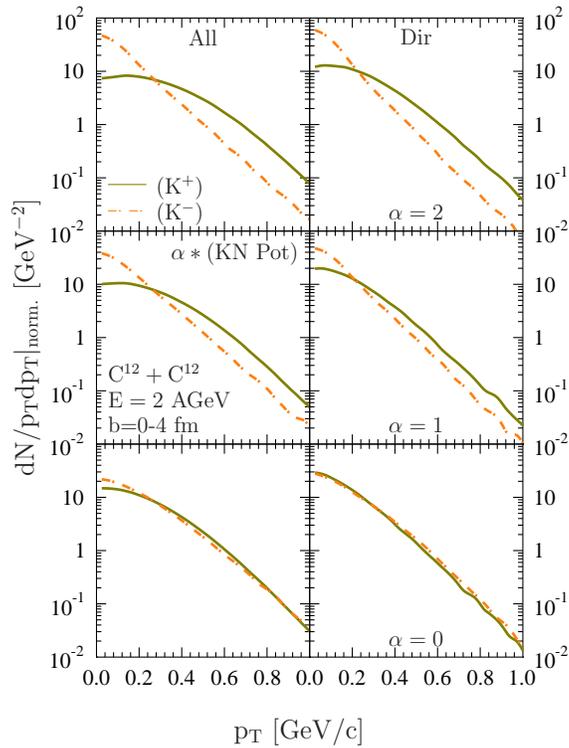,width=0.50\textwidth} \caption{Normalized $p_{T}$ spectra of kaons for different 
strengths of KN potential.} \label{fig1}
\end{figure}
where the sign of the vector term $\pm \Sigma_v^0$ is positive for $K^+$ and negative for $K^-$. 
This leads to different energies of $K^+$ and $K^-$ in the medium \cite{PR}.
The scalar potential $\Sigma_s$ is related to the $\sigma$ field which itself is related
in a non-linear way to the scalar
density  $\rho_s$. The vector potential is related to the baryon density $\rho_B$.
For details we refer to ref.\cite{Schaffner:1996kv}. Knowing $\omega({\bf{k}},\rho)$
we can calculate
\begin{equation}
\omega({\bf{k}}=0,\rho)
\end{equation}
which describes how the "mass" of the kaons changes if the meson is brought in a nuclear environment.

Korpa and Lutz \cite{Korpa:2004ae} have calculated $\omega({\bf{k}}=0,\rho)$ using a selfconsistent
Bethe Salpether equation.  The result of these calculations can be well approximated by
\begin{equation}
\omega({\bf{k}}=0,\rho)=m_{\rm K^+}(\rho )=m_{\rm K^+}(\rho =0) (1+ 
\alpha_{\rm K^+} \frac{\rho}{\rho_{0}}) \nonumber 
\end{equation}
\begin{equation}
 m_{\rm K^-}(\rho )=m_{\rm K^-}(\rho =0) (1+ 
\alpha_{\rm K^-} \frac{\rho}{\rho_{0}}) \label{kpmass}
\end{equation}
with $\alpha_{\rm K^+} = 0.07$ and $\alpha_{\rm K^-} = -0.22$. To study the influence of the potential we multiply
$\alpha_{K^+}$ and $\alpha_{K^-}$ by a artificial factor $\alpha$, as indicated in eq. \ref{schaf}.

The results which we present have been obtained with the IQMD program, an event generator which simulates heavy ion
reactions from the initial separation of projectile and target up to the final distribution
of fragments, nucleons and mesons. The details of IQMD program on how
strange particles are described in this approach have been extensively described
in ref. \cite{PR}. We employ a soft hadronic equation of state,
the $NN \to NN \to K^+$ cross section of Sibirtsev \cite{Sibirtsev:1995xb} and 
the $NN \to N\Delta \to K^+$ cross section of Tsushima \cite{Tsushima:1998jz,Tsushima:1994rj}.
For $K^-$, following Randrup and Ko \cite{Randrup:1980qd}, we apply an isospin factor to the
corresponding NN channel ($\sigma(N\Delta)=0.75\sigma(NN)=0.6mb(E-E_{thres})$, with energies
measured in GeV in the hyperon rest frame). The $Y\pi \to K^- N$ cross section can be 
obtained by detailed balance from the measured $ K^- N \to Y\pi$ cross section.
For the present study we simulate the $^{12}C+^{12}C$ reactions at an incident energy of 2 AGeV 
in the impact parameter range of 0-4 fm. 

In fig. \ref{fig1}, we display the transverse momentum ($p_{T}$) 
\begin{figure}
\psfig{figure=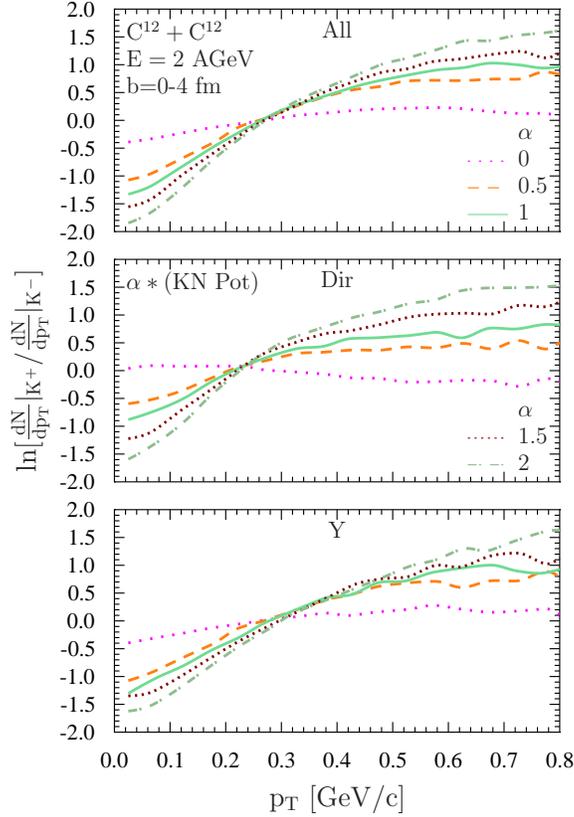,width=0.50\textwidth} 
\caption{Logarithmic ratio of $p_{T}$ spectra of $K^+$ and $K^-$ for different strengths of potential. We vary both the $K^+$N
and $K^-$N nucleus potential. Various lines are explained in the figure.}
\label{fig2}
\end{figure}
\begin{figure}
\psfig{figure=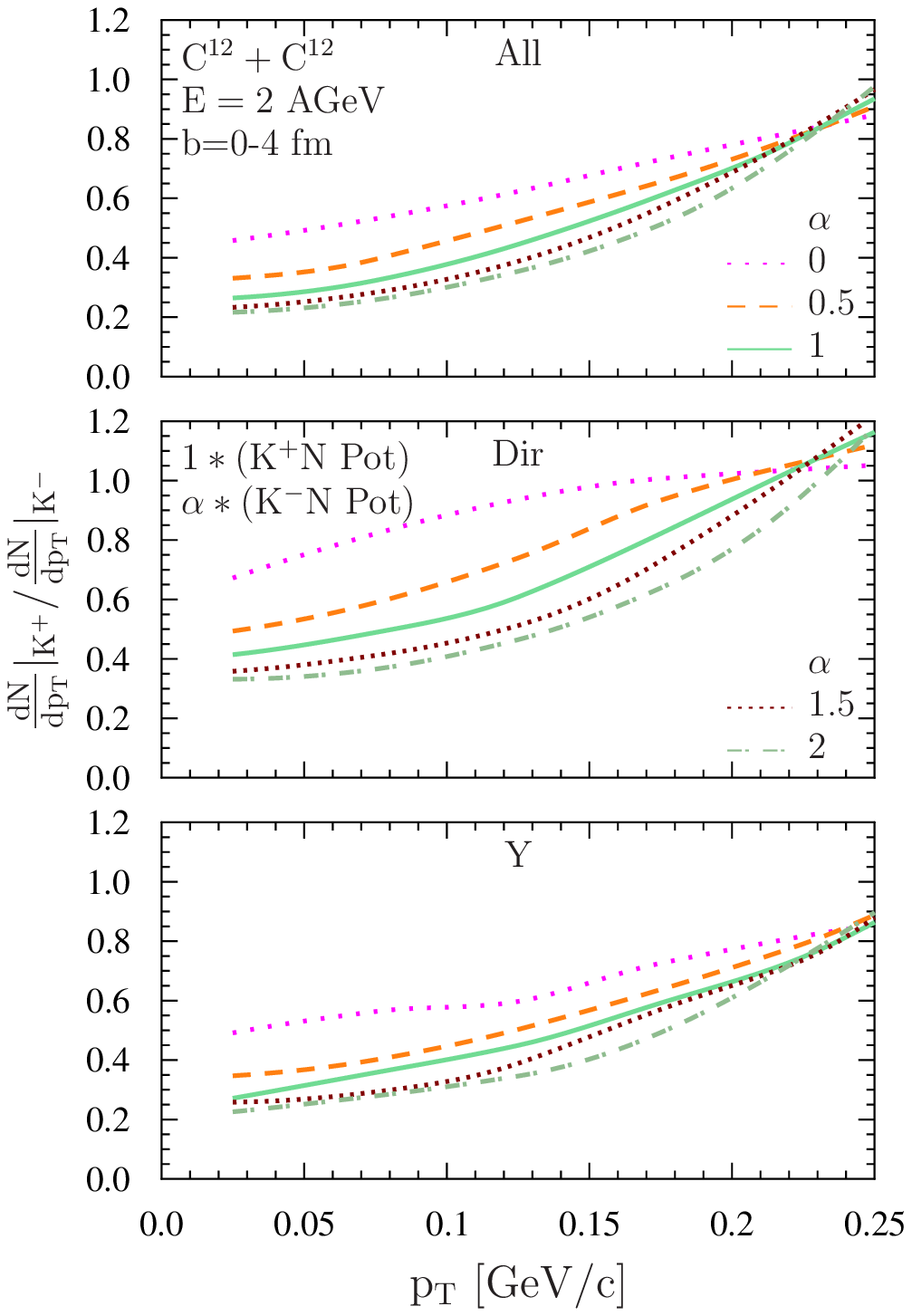,width=0.50\textwidth} 
\caption{Same as fig. \ref{fig2} but only $K^-$ nucleus potential is varied for a fixed $K^+$ nucleus potential. }
\label{fig3}
\end{figure}
the spectra of $K^+$ (solid) and $K^-$ (dash dotted line)  for $0\le b\le 4$ fm obtained for CC collisions at $E_{beam}=2 AGeV$.
The K potentials are used for both, $K^+$ and $K^-$,  and $\alpha$= 2,1,0  from the top to the bottom panels.  
The spectra is normalized to 1 so that the $K^+$ and $K^-$ spectra can directly be compared. Left and right panels represent the spectra for all the kaons (labelled as All)  and kaons produced in reaction BB $\to$ BB$K^+$$K^-$ (labelled as Dir), respectively. When we switch off the K nucleus potential ($\alpha$=0, bottom panel), the shape of the $K^+$ and $K^-$ spectra is almost identical.  We start the discussion with the right panel. For a vanishing potential 
the spectra are identical for both kaons. 
\begin{figure}
\psfig{figure=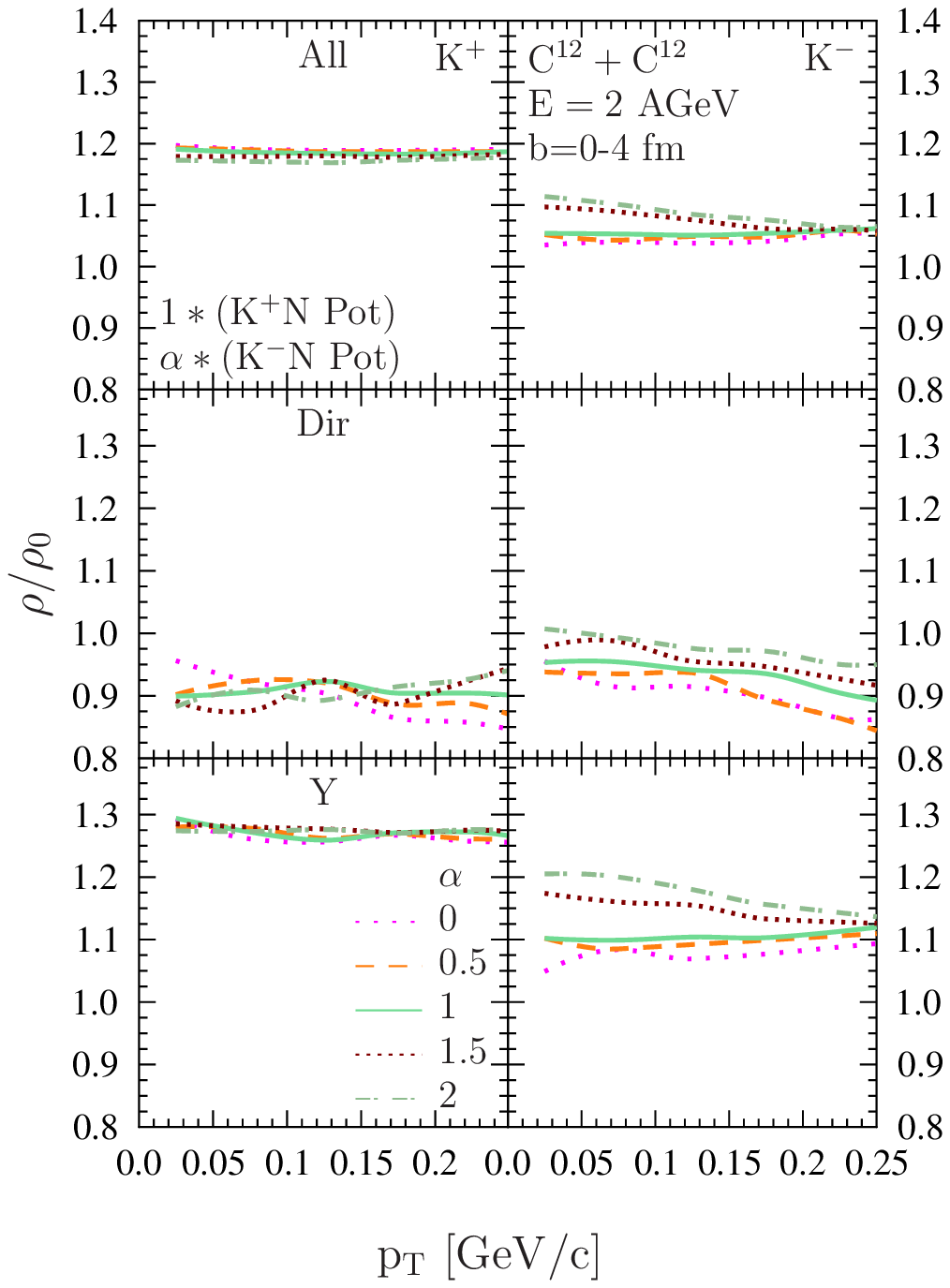,width=0.50\textwidth}\caption{Density at which the finally observed kaons are produced. On the
left (right) hand side we display the density distribution for the $K^+$ ($K^-$). The top panel shows the distribution for all events in which a $K^+$  and a $K^-$ is produced, the middle part for those events in which the $K^+$  and a $K^-$ are produced simultaneously and the bottom part for those events in which the $K^-$ is produced in a secondary collision.}
\label{fig4}
\end{figure}
The direct production is determined by the 4 body phase space and hence no difference
is expected in the production process. 
The fact that also finally the spectra are almost identical indicates that for these light systems rescattering has little influence on the spectral form. If we switch on  the potential we find $ \omega({\bf{k}}=0,\rho) < 500 MeV$ for the $K^-$ and $ \omega({\bf{k}}=0,\rho) > 500 MeV$ for the $K^+$. When the kaons leave the nucleus they have to get rid
of the excess mass ($K^+$) or they have to acquire mass ($K^-$).  For the $K^+$ a part of this excess mass is converted into kinetic energy whereas for the $K^-$ a part of the kinetic energy is converted into mass. (In IQMD  the total momentum and the total energy of all particles is conserved).  Consequently,  the number of $K^-$ ( $K^+$ ) increases (decreases) in the low momentum region causing the different slopes of the spectra for $K^-$ and $K^+$. This effect increases with 
increasing strength of potential (top panel). 

The left panels show that for vanishing  potential  the $K^-$ produced in a secondary collisions have almost the same 
slope as the directly produced. Increasing the potential we see that the effect which we have observed on the right hand
side survives if we include all $K^-$. This is essential because both classes can experimentally not be discriminated.
In fig. \ref{fig2} we display the logarithm of the ratio of the $p_{T}$ spectra of $K^+$ and $K^-$. Top, middle, and bottom panels show  this ratio for all $K^-$ ,  for the directly produced $K^-$ and for those produced in secondary collisions, respectively. Different lines are for different $\alpha(K^+)=  \alpha(K^-)$ values . The total yields depend on the choice of $\alpha$. 
The ratio is nearly constant without a K nucleus potential ($\alpha$=0, dotted magenta line). When we switch on the potential the ratio changes strongly in the low momentum region and decreases with increasing strength of the potential,
 whereas it remains nearly constant in the high momentum region. Comparing top and middle panel, we see that 
the influence of those $K-$ which come from secondary collisions on the spectral from at small $p_t$ is not essential.
This means that this ratio is almost exclusively sensitive to the potential and does not depend of the little or unknown cross sections.
 
Fig. \ref{fig3}  presents as well the ratio of the $K^+$ and $K^-$  spectra but this time the $K^+$ nucleus potential is taken as given by the theoretical predictions ($\alpha(K^+)$ =1) whereas for the $K^-$ we vary the potential assuming that the $K^+$ nucleus potentials can be determined by other means. This time we have chosen a linear scale.
We observe, as expected,  that the dependence of the slope on the $K^-$ nucleus potential becomes weaker as compared to a variation of both potentials but still varies by a factor of two and is hence a measurable quantity. 
This ratio depends on the  $K^-$ nucleus potential only and presents therefore the possibility to measure directly the strength  of the $K^-$ nucleus potential. It is therefore the desired 'smoking gun' signal to determine experimentally the $K^-$ potentials in matter at finite densities.  

It is interesting to see at which density the kaons are produced which are finally seen in the detector. This is displayed in
fig. \ref{fig4}. On the left (right) hand side we display as a function of $p_T$ the average density  at which those $K^+$ ($K^-$) are produced which are finally seen in the detectors. The top panel shows the density for all events in which a $K^+$  and a $K^-$ is produced, the middle part that for those events in which the $K^+$  and a $K^-$ are produced simultaneously and the bottom part for those events in which the $K^-$ is produced in a secondary collision.
Independent of the potential the kaons are produced at densities around normal nuclear matter density. The density 
for the directly produced kaons  is slightly lower than that of the other events because the higher the density the higher is also the probability that the $K^-$ is reabsorbed in a $\Lambda$.

In summary,   $K^+$ and $K^-$ spectra at  low transverse momentum, measured in  light symmetric systems
at around 2AGeV, depend strongly on the K nucleus potential. The ratio of the spectra allows therefore  for a direct determination of 
the strength of the $K^+$ as well as that of the $K^-$ potential  in a hadronic environment. The kaons are produced close to normal nuclear matter density and therefore this ratio is sensitive to the potential strength at that density. Other little known or unknown input quantities like the production or rescattering cross sections of $K^+$ and $K^-$  mesons do not spoil this signal.  Assuming that the $K^+$ spectra can be calculated reliably or is known from other sources we have shown that
this ratio will allow for a determination of the  $K^-$ nucleus potential strength and therefore contribute to solve the controversial discussion whether $K^-$ condensates can be expected in heavy ion collisions, whether kaonic cluster can be expected  and whether the potential strength, calculated for  heavy ion collision, is compatible with that observed in $K^-$ Atoms.

 Acknowledgement: We acknowlege the support of CEFIPRA under the contract 4101-1.

\

\end{document}